\documentclass[10pt,conference]{IEEEtran}
\IEEEoverridecommandlockouts
\pdfoutput=1
\usepackage{cite}
\usepackage{amsmath,amssymb,amsfonts}
\usepackage{algorithmic}
\usepackage{graphicx}
\usepackage{textcomp}
\usepackage{xcolor}

\usepackage[numbers]{natbib}
\usepackage{listings}
\usepackage{amsmath, amssymb}
\usepackage{makecell}
\usepackage{multirow}
\usepackage{booktabs}
\usepackage{subfigure}
\usepackage{url}
\usepackage{pifont}
\usepackage{threeparttable}

\usepackage[colorlinks,
linkcolor=blue,
anchorcolor=blue,
citecolor=blue]{hyperref}

\def\BibTeX{{\rm B\kern-.05em{\sc i\kern-.025em b}\kern-.08em
    T\kern-.1667em\lower.7ex\hbox{E}\kern-.125emX}}

\begin{document}
	
\title{BinMLM: Binary Authorship Verification with Flow-aware Mixture-of-Shared Language Model \\

\author{
	\IEEEauthorblockN{Qige Song$^{\ast\dag}$, Yongzheng Zhang$^{\dag}$, Linshu Ouyang$^{\ast\dag}$, Yige Chen$^{\ast\dag}$}
	\IEEEauthorblockA{$^\ast$ Institute of Information Engineering, Chinese Academy of Science, Beijing, China}
	\IEEEauthorblockA{$^\dag$ School of Cyber Security, University of Chinese Academy of Sciences, Beijing, China}
	\IEEEauthorblockA{songqige@iie.ac.cn, zhangyz@cacts.cn, ouyanglinshu@iie.ac.cn, chenyige@iie.ac.cn}
}
}

\maketitle

\begin{abstract}
Binary authorship analysis is a significant problem in many software engineering applications. In this paper, we formulate a binary authorship verification task to accurately reflect the real-world working process of software forensic experts. It aims to determine whether an anonymous binary is developed by a specific programmer with a small set of support samples, and the actual developer may not belong to the known candidate set but from the wild. 

We propose an effective binary authorship verification framework, BinMLM. BinMLM trains the RNN language model on consecutive opcode traces extracted from the control-flow-graph (CFG) to characterize the candidate developers' programming styles. We build a mixture-of-shared architecture with multiple shared encoders and author-specific gate layers, which can learn the developers' combination preferences of universal programming patterns and alleviate the problem of low training resources. Through an optimization pipeline of external pre-training, joint training, and fine-tuning, our framework can eliminate additional noise and accurately distill developers' unique styles.

Extensive experiments show that BinMLM achieves promising results on Google Code Jam (GCJ) and Codeforces datasets with different numbers of programmers and supporting samples. It significantly outperforms the baselines built on the state-of-the-art feature set (4.73\% to 19.46\% improvement) and remains robust in multi-author collaboration scenarios. Furthermore,  BinMLM can perform organization-level verification on a real-world APT malware dataset, which can provide valuable auxiliary information for exploring the group behind the APT attack.

\end{abstract}


\section{Introduction}
\subsection{Background and Motivation}
\begin{figure}[!t]
	\centering
	\includegraphics[width=6.6cm]{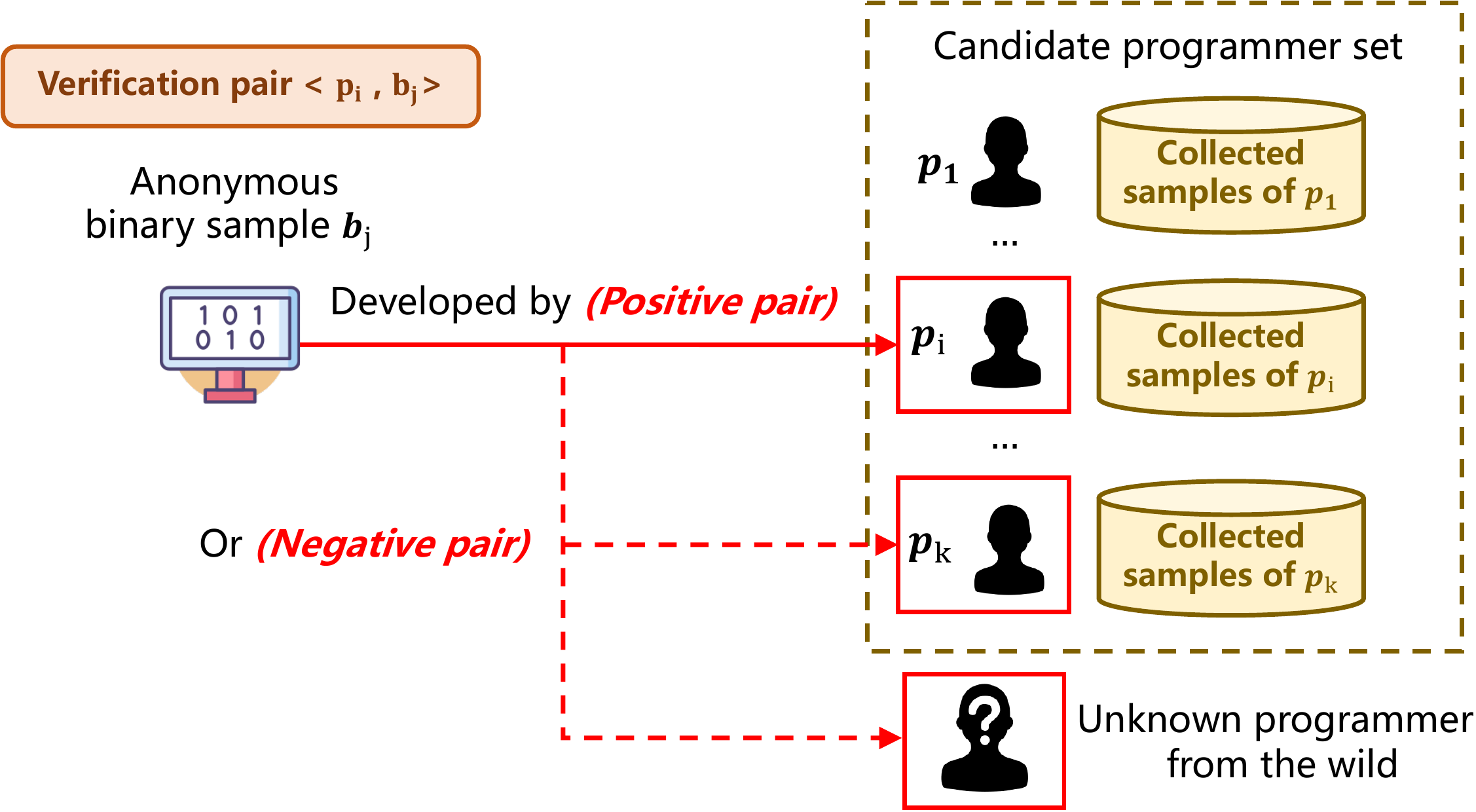}
	\caption{Binary authorship verification task}
	\label{intro_task}
	\vspace*{-1.2\baselineskip}
\end{figure}

Binary authorship analysis aims at recovering the developer’s identity information through executable binary files. It is critical in various software engineering applications, especially security-related scenarios, such as copyright infringement detection, cybercrime tracking, and malicious software (malware) forensics, where source code is rarely available. In this paper, we formulate the problem as a practical \emph{binary authorship verification task}. As shown in figure \ref{intro_task}, given an anonymous binary sample, our goal is to determine whether it is developed by a specific programmer of the candidate set. Each candidate programmer has a small set of binaries that can characterize his programming style. The difficulty of the task is that the actual developer of the questioned binary is likely not to belong to the candidate set but from the wild, which is in line with the realistic experiences of professional binary analysts \cite{marquis2015big}.

Binary authorship verification is nontrivial as it faces two main challenges: 

\textbf{(a)} Binary samples of the same programmer may implement relevant or completely different functions. Prior researchers manually designed programming stylistic features to establish the unified programmer templates. It heavily relies on expert experience and can not capture specific programming behaviors that have not appeared in the known samples. 

Deep learning can perform automatic feature extraction and has strong generalization capabilities. It has achieved promising results in many software engineering applications \cite{wang2020detecting} \cite{tian2020evaluating} \cite{wang2021mulcode} \cite{bui2021infercode}. However, the collected samples of candidate authors are usually very limited, making it challenging to train the deep neural network with a large number of parameters.

\textbf{(b)} Modern software development is usually completed by cooperation, even including malware production, like APT attackers usually cooperate to launch staged events. The collaborators' code snippets may introduce noise when identifying the major contributor, and it is more challenging to implement organization-level authorship verification due to the mixture of different programmers' styles.

\subsection{Limitation of Prior Art}
Most previous binary authorship analysis approaches try to solve the problem by traditional supervised machine learning algorithms \cite{rosenblum2011wrote} \cite{alrabaee2014oba2} \cite{meng2016fine} \cite{alrabaee2018leveraging} \cite{caliskancoding}. They extract the known programmers' stylistic features from training examples and identify the developer of the anonymous binaries in the test stage. Table \ref{qualitative_sota_table} lists the state-of-the-art approaches. Rosenblum. et al. \cite{rosenblum2011wrote} extracted bytecode n-gram, instruction idioms, and CFG graphlets to create author-style templates. BinAuthor \cite{alrabaee2018leveraging} constructed the authors' choices set to recognize their programming habits. Caliskan-Islam. et al. \cite{caliskancoding} extracted hybrid features from disassembly instructions and decompiled code to model the author's style survives compilation. 

The above approaches require manual feature engineering, which relies on domain knowledge and has been shown to be dataset-specific. To automatically extract the programmer's style characteristics, Bineye \cite{alrabaee2019bineye} proposed a deep learning-based authorship attribution method, which uses three convolutional neural networks (CNNs) on gray-scale images, opcode sequences, and function invocations. Although showing performance improvement, deep neural networks rely on sufficient collected samples with annotations to fit the large-scale parameters, which is usually unrealistic in real-world binary forensics scenarios.

Furthermore, one common dilemma of most existing approaches is the inability to handle the developers who do not belong to the candidate set. Caliskan-Islam \emph{et al.} \cite{caliskancoding} decided to accept or reject the results according to the classifier's confidence score. Rosenblum \emph{et al.} \cite{rosenblum2011wrote} calculated the stylistic similarities between anonymous binary pairs based on a distance metric for authorship clustering. However, their performance is not ideal when dealing with large size of candidates and insufficient samples. In this paper, we consider the real-life experience of professional binary analysts and formulate the \emph{binary authorship verification task}. Our goal is to determine whether a binary is developed by a candidate author with limited samples based on the preserved programming style.

\begin{table}
	\caption{State-of-the-art binary authorship analysis approaches}
	\label{qualitative_sota_table}
	\centering
	\begin{tabular}{m{84pt}<{\centering}|m{37pt}<{\centering}|m{94pt}<{\centering}}
		\toprule
		Approach & Feature engineering & Handling unknown programmers from the wild \\
		
		\midrule
		Rosenblum \emph{et al.} \cite{rosenblum2011wrote} & Manual & \ding{51}  (by stylistic similarity) \\
		Oba2 \cite{alrabaee2014oba2} & Manual & \ding{55} \\
		Meng \emph{et al.} \cite{meng2016fine} & Manual & \ding{55} \\
		BinAuthor \cite{alrabaee2018leveraging} & Manual & \ding{55} \\
		Caliskan-Islam \emph{et al.} \cite{caliskancoding} & Manual  & \ding{51}  (by confidence score) \\
		Bineye \cite{alrabaee2019bineye} & Automatic & \ding{55} \\
		\bottomrule
	\end{tabular}
	\vspace*{-1.2\baselineskip}
\end{table}

\subsection{Proposed Method}
In this paper, we propose \textbf{BinMLM}, a \underline{bin}ary authorship verification framework with flow-aware \underline{m}ixture-of-shared \underline{l}anguage \underline{m}odel. It addresses the above challenges as follows:

Specific programming languages have strict syntax restrictions. Therefore, developers' personal habits only account for a small percentage of the program, while most parts are overlapping general patterns. We interpret programmers' personal style as the selection and combination of \emph{universal programming patterns}, such as the preferences for specific programming paradigms, branch and loop forms, organization structures of user-defined functions, and exception handling ways. These stylistic preferences are implicit in the instruction execution sequence.

We train RNN language models with powerful sequence modeling capabilities on the disassembly code to automatically characterize the styles of candidate programmers, avoiding explicit feature engineering. To recover the high-level stylistic patterns from the binaries, we extract consecutive opcode traces as the essential units of the language model and preserve the flow information in the basic block bi-grams of CFG.

We introduce a \emph{mixture-of-shared} architecture with multiple encoders shared among all developers to model the universal programming patterns from different views. Each programmer is assigned a gate layer to learn the specific combination weights of shared representations, reflecting their preferences of multiple generic patterns. Compared to training a separate language model for each author, this layer sharing mechanism can alleviate the insufficient data problem by effectively utilizing the small-scale training samples.

We utilize an \emph{optimization pipeline} to reduce the noise of collaborators and further isolate the small proportion of unique styles from generic programming patterns. We first train the overall parameters jointly. Specifically, we randomly select an author's mini-batches each time and optimize the shared encoders and the corresponding author-specific layers. After that, we fix the parameters of shared layers and separately fine-tune each author-specific decoder with specially designed regularization terms. Joint training enables the model to learn the general programming patterns better. Author-specific fine-tuning facilitates the model to explore how to decode personal styles accurately, improving the robustness of major author characterization in multi-programmer collaboration scenarios.

\subsection{Key Contributions}
We summarize our major contributions as follows:

\begin{itemize}
	\item We formulate a practical \emph{binary authorship verification} task, which considers that the developers of anonymous binaries may not belong to the candidate set but from the wild. Our task setting is in line with the working process of software forensic experts, and we provide an effective solution for them to perform automatic analysis.
	
	\item We creatively characterize the authors' programming styles by training the RNN language model on flow-aware instruction execution traces, which can avoid manual feature engineering. 
	
	\item We design a novel \emph{mixture-of-shared} language model and an effective \emph{optimization pipeline}. With multiple shared encoders and author-specific gate layers, we can fully utilize limited samples and model the developers' combination preferences of general programming patterns. Furthermore, our \emph{optimization pipeline} can eliminate additional noise and accurately distill developers' unique stylistic characteristics.
	
	\item 
	We conduct extensive experiments to evaluate BinMLM on datasets with different numbers of developers and samples. Results show that it can surpass baselines built on the state-of-the-art feature set by a large margin (AUC = 0.83 $\sim$ 0.94, 4.73\% $\sim$ 19.46\% improvement). Moreover, BinMLM remains robust in multi-programmer collaboration scenarios and can perform practical organization-level verification on a real-world APT malware dataset.
	
\end{itemize}

\section{Problem Defination}
In this section, we formalize the binary authorship verification task as follows:

We first construct a candidate set $P$ = ($p_1$, $p_2$, ..., $p_n$) of $n$ known programmers, and each programmer $p_i$ has several previously collected binary samples $B_i$ = ($b_{i_1}$, $b_{i_2}$, ... ,$b_{i_m}$). Then for an authorship verification pair $\langle p_i, b_j \rangle$, we aim to determine whether the anonymous binary $b_j$ is developed by the candidate programmer $p_i$. Note that we assume $b_j$  is developed by a single programmer or has a major contributor in multi-programmer collaboration scenarios, and the actual developer of $b_j$ may belong to $P$ or may come from the wild.

Previous binary authorship analysis methods directly adopt a close-world classification setting, which assumes the anonymous binaries to be analyzed are all developed by known programmers from the candidate set. It is inconsistent with the real-world software forensic scenarios, and our proposed binary authorship verification task can handle the problem from a more realistic and comprehensive perspective.

\section{Preliminaries}
\subsection{RNN Language Model}
Language models has shown strong text style characterization abilities in previous literature \cite{bagnall2015author} \cite{ge2016authorship} \cite{ouyang2020gated}. For an  input sentence $S$ = ($u_1$, $u_2$,..., $u_l$), language model quantitatively describe the joint probability $p(S)$ as:

\begin{equation}
	p(S)=\prod_{i=1}^{l} p\left(u_{t} \mid u_{1:t-1}\right)
\end{equation}
$u_i$ denotes the statistical unit of the language model. In this paper, we train the language model on the disassembly instruction sequence to characterize developers' programming styles. We use recurrent neural network (RNN) to encode $u_{1:t-1}$ \cite{mikolov2010recurrent}. For an input sequence, we first transform each token $u_i$ into a vector $\mathbf{e}_t$ through an embedding layer, then feed the sequence into RNN layer to model the context representation. At each time-step $t$ of RNN, the hidden state $\mathbf{h}_t$ is updated as follows:

\vspace*{-0.3\baselineskip}
\begin{equation}
	\mathbf{h}_t=RNN\left(\mathbf{h}_{t-1}, \mathbf{e}_{t-1}\right)
\end{equation}

After that, $\mathbf{h}_t$ is fed into a fully-connected layer with \emph{softmax} function, which acts as a decoder to estimate the probability distribution of the next unit:

\vspace*{-0.3\baselineskip}
\begin{equation}
	y\left(u_{t}\right)=f\left(\mathbf{W} \cdot \mathbf{h}_t+\mathbf{b}\right)
\end{equation}

\vspace*{-0.8\baselineskip}
\begin{equation}
p\left({u_t} \mid u_{1:t-1}\right)=\frac{e^{ \left(y\left(u_{t}\right)\right)}}{\sum_{k=1}^{|V|} e^{ \left(y\left(u_{k}\right)\right)}}
\end{equation}
$f$ is the nonlinear activation function, $|V|$ is the vocabulary size of the statistical units. The loss function of the language model is defined as the cross-entropy between the predicted distribution and the ground-truth:

\vspace*{-0.3\baselineskip}
\begin{equation}
	L(S)=-\sum_{t=1}^{m} u_t \log \left(p\left({u_t} \mid u_{1:t-1}\right)\right)
	\vspace*{-0.3\baselineskip}
\end{equation}

\begin{figure}[!t]
	\centering
	\includegraphics[width=6cm]{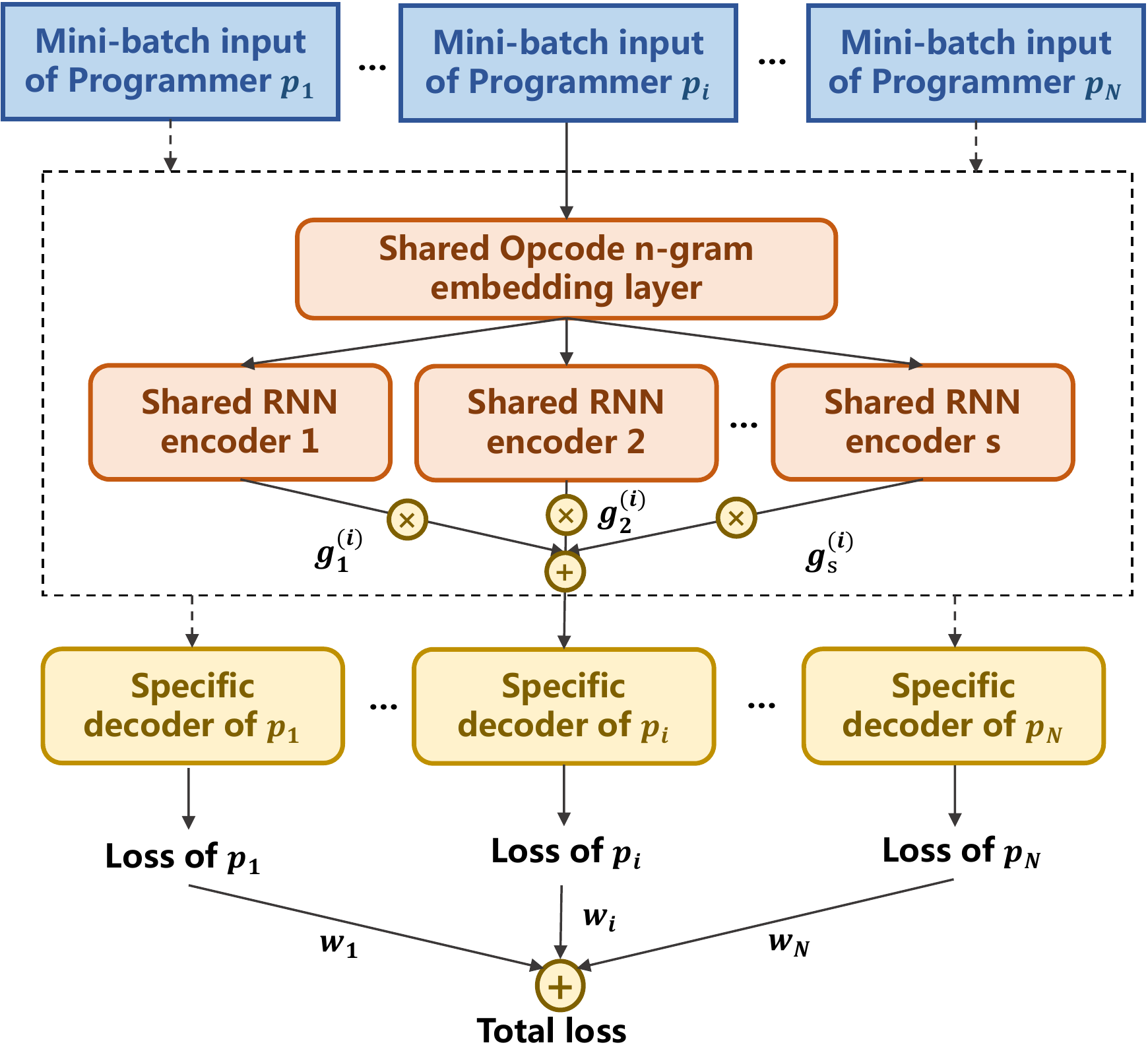}
	\caption{Mixture-of-Shared language model architecture}
	\label{mos_arch}
\vspace*{-1.2\baselineskip}
\end{figure}

\begin{figure*}[!t]
	\centering
	\includegraphics[width=13cm]{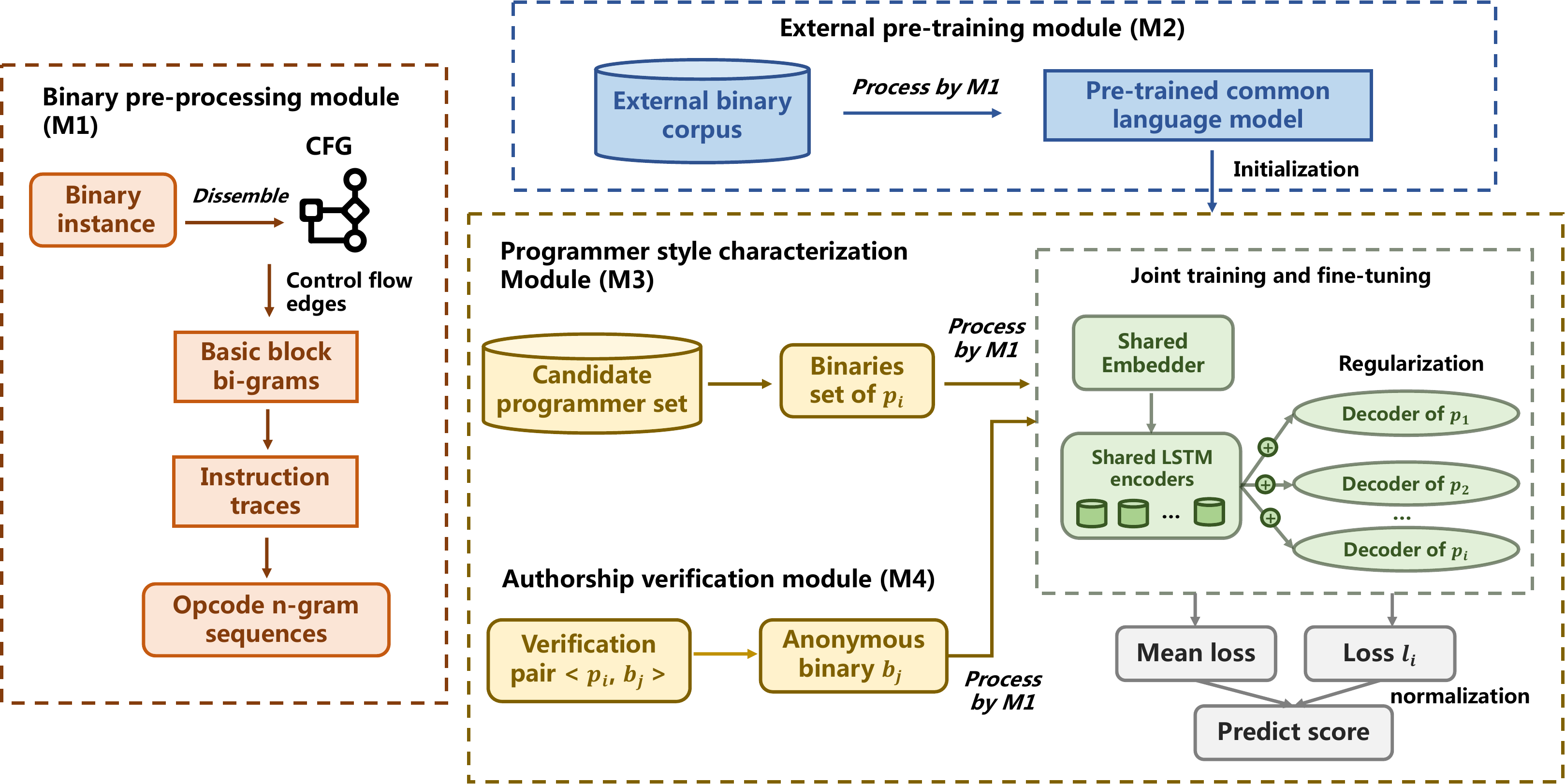}
	\caption{Binary authorship verification workflow of BinMLM}
	\label{sys_overview}
	\vspace*{-1.2\baselineskip}
\end{figure*}

\subsection{Mixture-of-Shared Network Architecture}
To conduct binary authorship verification formalized in section \uppercase\expandafter{\romannumeral2}, we expect to train a language model for each candidate author on his binary samples. Each author's programming style will lead to the differences of the learned probability $p(S)$, and a binary has a relatively high probability on its actual developer's language model. Since the number of candidate programmers can be large, setting an independent network for each author will limit the model's scalability. In addition, the syntax rules of programming languages are much stricter than natural language. Developers use a fixed set of keywords and APIs, and the assembly instructions compiled from high-level source code have more limited operations types. Therefore, developers' personal styles mainly come from the selection and combination of universal programming patterns.

Based on these considerations, we exploit a mixture-of-shared network to train the language models of candidate programmers simultaneously. Figure \ref{mos_arch} shows the network architecture. We set up multiple RNN encoders shared among all programmers to map the processed samples into different subspaces, modeling the generic programming patterns from different perspectives. For programmer $p_i$, the shared encoder $j$ will generate the corresponding hidden representation $\mathbf{h}_{j}^{(i)}$.

To extract the developer's specific preference for the combination of common programming patterns, we build a separate gate layer for each programmer to learn the importance weights of shared representations generated by different RNN encoders. Specifically, the weight parameters $\mathbf{g_{i}}$ of programmer $p_i$ is calculated by a feed-forward layer as follows:

\vspace*{-0.3\baselineskip}
\begin{equation}
	\mathbf{g}_{i}=\sigma \left(\mathbf{W}_{g_i} \cdot  \mathbf{E}_{i}+\mathbf{b}_{g_i}\right)
\end{equation}
$\mathbf{E}_{i}$ denotes the embedded vectors of the input sequence. We compute the context representation $\mathbf{r}_{i}$ by the element-wise multiplication of $\mathbf{h}_{j}^{(i)}$ with their corresponding gates weights:

\vspace*{-0.3\baselineskip}
\begin{equation}
	\mathbf{r}_{i}=\sum_{j=1}^{s} \mathbf{g}_i \cdot \mathbf{h}_{j}^{(i)}
\end{equation}
$s$ is the number of shared RNN encoders. $\mathbf{r}_{i}$ is finally fed into the corresponding author-specific linear decoder to generate predictions of subsequent units.

Our model is inspired by the text style characterization approaches proposed by Bagnall \cite{bagnall2015author} and Ouyang \emph{et al.} \cite{ouyang2020gated}, and the multi-task learning network designed by Ma \emph{et al.} \cite{ma2018modeling}. Such an architecture allows all programmers' samples to optimize the shared layers, and in turn optimize each specific decoder, improving the generalization ability of low-resource programmers' models. Moreover, multiple shared encoders and author-specific gates can flexibly combine the subspace of the universal programming patterns, accurately characterizing the programmer's personal style.

\section{Proposed Method}
\subsection{Overview}
The general idea of our proposed binary authorship verification framework BinMLM is to characterize the programmer's unique style contained in binary files by language model and simultaneously learn specific language models of large numbers of candidate programmers. To this end, BinMLM involves four major modules (shown in Figure \ref{sys_overview}).

\begin{itemize}
	\item Binary pre-processing module (\emph{Module 1}): We disassemble the input binary file and extract the CFG structure. Then we extract the instructions traces with control flow semantics and use the opcode n-grams as the essential units of the subsequent programming style characterization process. This module is a fundamental component shared by the following modules.

	\item External pre-training module (\emph{Module 2}): We construct a large-scale external binary code corpus without authorship annotations and pre-process it by \emph{Module 1}. Then we pre-train a general language model on these wild binary files to initialize the subsequent models.
		
	\item Candidate programmer style characterization module  (\emph{Module 3}): We introduce a mixture-of-shared architecture with multiple shared encoders and author-specific gate layers and decoders to train the language models of all candidate programmers. We first jointly optimize the overall parameters by binary samples of all programmers, then freeze the shared layer and individually fine-tune the separate decoders with additional regularization items.
	
	\item Authorship verification module  (\emph{Module 4}): For a verification pair $\langle p_i, b_j \rangle$. We pre-process binary $b_j$ and evaluate its loss value $l_i$ on the trained language model of programmer $p_i$. Then we use the loss array $\mathbf{L}^{(j)}$ of $b_j$ on all candidate programmers' language models to normalize $l_i^{(j)}$ and obtain the final verification score.
 
\end{itemize}

\subsection{Binary File Pre-processing}
\begin{figure*}[!t]
	\centering
	\includegraphics[width=12cm]{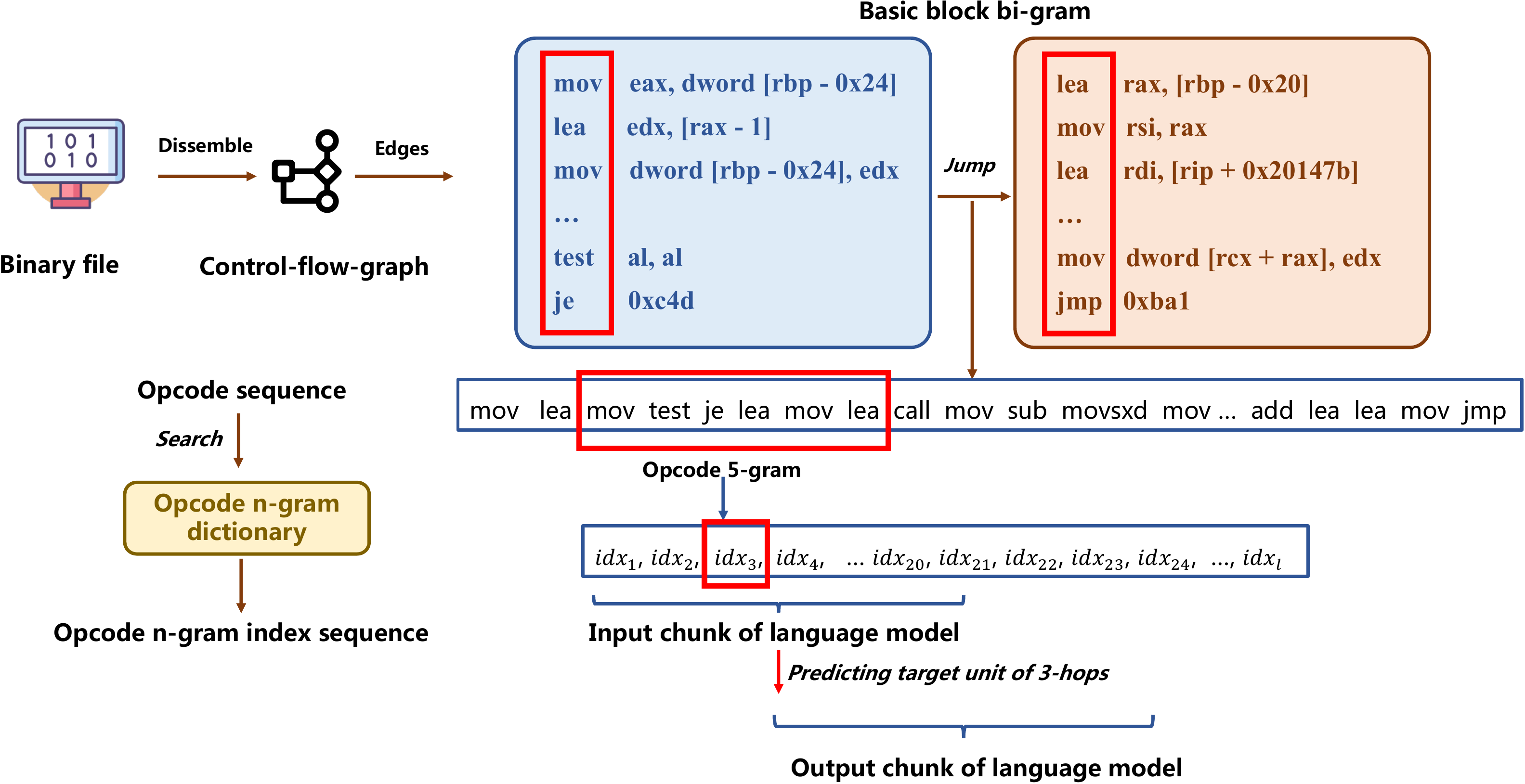}
	\caption{Binary file pre-processing module}
	\label{binary_process}
\vspace*{-1.2\baselineskip}
\end{figure*}

Figure \ref{binary_process} shows the details of our binary file pre-processing module. The major components include processing the disassembly code and building language models on the collected binary samples.

\subsubsection{Process disassembly code}
A binary to be analyzed is represented as discrete byte streams, and we use disassembling tool to extract its corresponding CFG structure. The nodes in CFG, called basic blocks, act as the smallest units of sequential instruction executions. The edges represent the control flow transfer between basic blocks, which may be caused by \emph{if-else} branches, loop structures, or jumps across blocks.

To preserve the flow transfer semantics in CFG, we first connect the endpoint basic blocks of all edges as basic block bi-grams. Then we extract the instruction traces within the two basic blocks sequentially as the sequence $I$ = ($i_1$, $i_2$, ..., $i_l$), $l$ is the number of instructions. Assembly instruction is composed of an opcode and zero to more operands. The opcode specifies the operation to be conducted. The operands specify registers, immediate literals, or memory locations, which can be different under various compilation environments. To improve the versatility of our framework and reduce the scale of the language model's vocabulary, we only keep the opcode in the instruction sequence and update the sequence to $O$ = ($o_1$, $o_2$, ..., $o_l$), $o_i$ represents the opcode at position $i$.

\begin{figure}[!t]
	\centering
	\includegraphics[width=9cm]{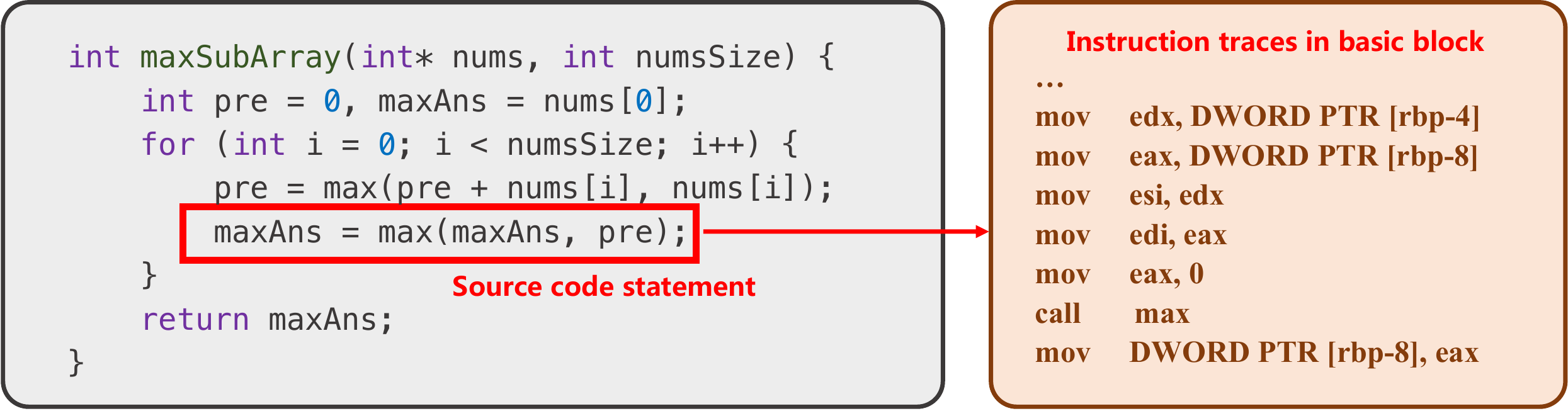}
	\caption{Source code statements and counterpart assembly instructions}
	\label{source_assembly}
	\vspace*{-1.2\baselineskip}
\end{figure}

We explore how to better characterize the developer’s programming style in the binaries by comparing the disassembly code with the original high-level source code. Figure \ref{source_assembly} shows a source code snippet and part of its corresponding assembly instruction traces. We can see that a short statement in the source code, such as $maxAns=max(maxAns, pre)$, will be transformed into a relatively long assembly instruction sequence, and the corresponding opcode sequence is $MOV$, ..., $CALL$, $MOV$, which implements the operations of register loading, parameter reading, and function invocations. Therefore, consecutive short opcode traces can potentially reflect the programming patterns of source code statements. Since the number of the assembly instruction complied from each statement is uncertain, we slide a window fixed to size $n$ on the original opcode sequence to obtain $G$ = ($g_1$, $g_2$, ..., $g_l$), $g_i$ denotes an opcode n-gram representing a small continuous sequence of operations composed of $n$ instructions.

\subsubsection{Build language model for binaries}
We train the RNN language model on the extracted opcode n-gram sequences of binaries to characterize the developer's programming style. The RNN encoding layer extracts the context representation of the previous opcode n-gram units, and the linear decoding layer estimates the probability distribution of the next unit. In our experiment, we found that setting the subsequent opcode n-gram unit of multiple hops as the prediction target of the language model can generate more distinguishable author-specific styles. Because increasing the task difficulty encourages the model to dig out more robust programming patterns from the training samples.

\subsection{External Language Model Pre-training}
The collected binaries of candidate programmers are insufficient for training RNN language models with a large number of parameters. Note that traditional text language models are usually trained on millions of tokens, but our collected samples may only contain a few thousand opcode n-gram units. To tackle this problem, we first pre-train a common RNN language model on the opcode n-gram sequences extracted from a large-scale disassembled external binary corpus. The pre-trained model, with an opcode n-gram embedding layer, an LSTM encoding layer, and a linear decoding layer, contains the general programming features of these wild binary files. It will be used to initialize the parameters of candidate programmers' language models.

\subsection{Candidate Programmer Style Characterization}
\subsubsection{Mixture-of-shared language model}
We use the collected binaries to train an author-specific language model for each candidate programmer. As defined in section \uppercase\expandafter{\romannumeral2}, there are $n$ programmers in the candidate set $P$ = ($p_1$, $p_2$, ..., $p_n$). Each programmer $p_i$ has a set of binary samples $B_i$ = ($b_{i_1}$, $b_{i_2}$, ... , $b_{i_m}$). We process each binary in $B_i$ as control-flow aware opcode n-gram sequences, and each sequence is represented as $G_i$ = ($g_{i_1}$, $g_{i_2}$,..., $g_{i_l}$). The processed sequences of the programmers' binary samples are grouped into input mini-batches to train the corresponding language models.

We utilize the mixture-of-shared architecture elaborated in section \uppercase\expandafter{\romannumeral3} to train the binary code language models of all candidate programmers. We first deploy a shared opcode n-gram embedding layer to convert the input sequence $G_i$ into a vector sequence $\mathbf{E_i}$ = ($\mathbf{e}_{i_1}$, $\mathbf{e}_{i_2}$,  …, $\mathbf{e}_{i_l}$). Then we use multiple shared RNN encoders with the same structure to encode common programming patterns from different views. The $j$-th shared RNN encodes the serialized context information into hidden state vectors $\mathbf{H_i}$ =  ($\mathbf{h}_{i_1}^{(j)}$, $\mathbf{h}_{i_2}^{(j)}$,  …, $\mathbf{h}_{i_l}^{(j)}$).  Next, for each developer, we set up a separate gate layer to produce a weighted aggregation of hidden states generated by multiple RNNs as Equation 7. It can learn the developers' combination preferences of universal programming patterns and form their personal style on this basis. Finally, we feed the mixed context representation $\mathbf{R_i}$ = ($\mathbf{r}_{i_1}$, $\mathbf{r}_{i_2}$,  …, $\mathbf{r}_{i_l}$) into $p_i$ 's independent linear decoder with the \emph{softmax} function to estimate the subsequent units. The author-specific decoder is trained on the developer's own samples.

We initialize the parameters of all shared layers and author-specific decoders by the language model pre-trained on the external binary code corpus. The loss $\operatorname{L}_{p_i}$ of programmer $p_i$'s language model is the average cross-entropy losses on the mini-batches constructed by his binary samples. The overall loss of our mixture-of-shared language model is the weighted sum of all programmers' losses, $w_i$ denotes the weight of $\operatorname{L}_{p_i}$:

\vspace*{-0.3\baselineskip}
\begin{equation}
	\operatorname{L}_{lm} = \sum_{i=1}^{n} w_i \cdot \operatorname{L}_{p_i}
\end{equation}

\subsubsection{Optimization pipeline}
The optimization pipeline of the mixture-of-shared language model is divided into a joint training phase and a fine-tuning phase to learn generic programming patterns and author-specific stylistic features. During the joint training phase, we randomly select a programmer's mini-batches as the input of each iteration, optimizing the shared layers and the corresponding specific gate layers and linear decoders. Then in the fine-tuning phase, we freeze the optimized shared embedding layer and LSTM encoders, as well as the gate layers, and individually fine-tune the decoders of each programmer. 

According to our analysis, the programmer's distinguishable unique styles only account for a small percentage of the binary code, while most parts are common programming patterns. Our optimization pipeline can effectively exploit the limited training samples of each candidate to distill the author-specific stylistic features and eliminate additional noise.

\subsubsection{Regularization of specific-decoders} 
Extracting opcode n-grams will cause the vocabulary of the statistical units to increase significantly as the value of $n$ increases, thereby enlarging the sample space of the language model. In the fine-tuning stage, each programmer's decoder is only optimized by its own small size of binary samples. To limit the parameter space of the author-specific language models, we add regularization terms between a pair of decoders to encourage their parameters to be more similar and prevent overfitting on the inadequate training data.

Specifically, when fine-tuning the linear decoder of programmer $p_i$, we calculate the $l_1$ distance among its parameters $\mathbf{W}_{p_i}$, $\mathbf{b}_{p_i}$, and the parameters of other programmer's decoders as the regularization loss. We average the regularization loss of all programmers as $\operatorname{L_{reg}}$ and add it to the loss of the overall language model $\operatorname{L_{lm}}$. $\lambda$ is the weight coefficient of $\operatorname{L_{reg}}$:

\vspace*{-0.3\baselineskip}
\begin{equation}
\operatorname{L_{reg}}=\sum_{i=1}^{n} \sum_{j=1, j \neq i}^{n}\left|\mathbf{W}_{p_i}-\mathbf{W}_{p_j}\right|+\left|\mathbf{b}_{p_i}-\mathbf{b}_{p_j}\right|
\end{equation}

\vspace*{-0.3\baselineskip}
\begin{equation}
\operatorname{L}_{total} = \operatorname{L}_{lm} + \lambda  \cdot \operatorname{L_{reg}}
\vspace*{-0.3\baselineskip}
\end{equation}

\subsection{Binary Authorship Verification}
In the binary authorship verification phase, for a verification pair $\langle p_i, b_j \rangle$, our goal is to determine whether the anonymous binary sample $b_j$ is developed by programmer $p_i$. We first process $b_j$ in the same way as the binary files of the training phrase, disassemble it into a CFG structure and extract the opcode n-gram sequences of the basic block bi-grams. Then we feed the processed sequences into the trained mixture-of-shared language models to get the corresponding loss array $\mathbf{L}^{(j)}$ = [${l_1}^{(j)}$, ${l_2}^{(j)}$, ..., ${l_i}^{(j)}$, ... ${l_n}^{(j)}$] of $b_j$ on the binary code language models of all programmers in the candidate set $P$. We jointly determine the verification result based on the loss value ${l_i}^{(j)}$ of $b_j$ on $p_i$'s language model and the loss array $\mathbf{L}^{(j)}$. If ${l_i}^{(j)}$ is relatively small, it is more likely that $b_j$ was developed by the programmer $p_i$.
 
To be more specific, we calculate the average value $\operatorname{Avg}({\mathbf{L}^{(j)})}$ of $\mathbf{L}^{(j)}$ and its variance $\operatorname{Var}({\mathbf{L}^{(j)})}$  to normalize ${l_i}^{(j)}$ and obtain the verification score $s(i, j)$ of pair $\langle p_i, b_j \rangle$:

\vspace*{-0.3\baselineskip}
\begin{equation}
s(i, j)=\frac{{l_i}^{(j)}-\operatorname{Avg}({\mathbf{L}^{(j)})}}{\operatorname{Var}({\mathbf{L}^{(j)}})}
\vspace*{-0.3\baselineskip}
\end{equation}

\begin{table*}
	\caption{Main results on different datasets}
	\begin{threeparttable}
		\label{baseline_res}
		\centering
		\begin{tabular}{p{40pt}<{\centering}p{125pt}<{\centering}|p{40pt}<{\centering}p{40pt}<{\centering}p{40pt}<{\centering}p{40pt}<{\centering}p{40pt}<{\centering}p{40pt}<{\centering}}
			
			\multirow{2}{*}{Approach} & \multirow{2}{*}{Setting} & \multicolumn{2}{c}{GCJ-C (N = 50, m = 5)} & \multicolumn{2}{c}{GCJ-C++ (N = 100, m = 5)} & \multicolumn{2}{c}{Codeforces (N = 100, m = 20)}  \\
			
			& & AUC-ROC & AP & AUC-ROC & AP & AUC-ROC & AP  \\
			\midrule
			
			\multirow{4}{*}{\texttt{Sim-Base}} & with full features & 0.8448 & \underline{0.8656} & \underline{0.7820} & 0.7950  & 0.7154 & 0.7209 \\
			& w/o CFG n-grams & \underline{0.8526} & 0.8526 & 0.7497 & 0.7596 & \underline{0.7159} & \underline{0.7267} \\
			
			& with opcode n-grams & 0.7587 & 0.7915 & 0.7718 & 0.7882 & 0.6132 & 0.6186 \\
			& with opcode n-grams + CFG n-grams & 0.7801 & 0.7979 & 0.7801 & \underline{0.7979} & 0.6239 & 0.6264 \\
			
			\midrule
			\multirow{2}{*}{\texttt{Con-Base}} & with feature selection & 0.7702 & 0.7719 & 0.6011 & 0.6014  & 0.6184 & 0.6155 \\
			& w/o feature selection & 0.7623 & 0.6249 & 0.6120 & 0.6153 & 0.7613 & 0.6209 \\

			\midrule
			 \multicolumn{2}{c|}{\textbf{BinMLM}} & \textbf{0.8929} & \textbf{0.9135} & \textbf{0.8414} & \textbf{0.8588} & \textbf{0.8552} & \textbf{0.8640} \\
			 
			 \multicolumn{2}{c|}{$\varDelta$ to the best results of baselines} & + 4.73\% & + 5.53\% & + 7.59\% & + 7.63\% & + 19.46\% & + 18.89\% \\
			 
			\bottomrule
		\end{tabular}
		
		\begin{tablenotes}
			\footnotesize
			\item[*]  GCJ is the abbreviation of Google Code Jam dataset. $N$ denotes the number of candidate programmers, $m$ is the number of support samples. The underlines indicate the best results of baselines. Bold indicates the overall best results.
		\end{tablenotes}
	\end{threeparttable}
	
	\label{main_result_table}
\end{table*}

\section{Evaluation}
In this section, we conduct extensive experiments to evaluate our proposed binary authorship verification framework, BinMLM. First, we describe the dataset and evaluation metrics of our experiments (section \uppercase\expandafter{\romannumeral5}.A). Next, we compare BinMLM with baselines built on the state-of-the-art feature set (section \uppercase\expandafter{\romannumeral5}.B.1) and evaluate BinMLM on datasets of different numbers of candidate programmers (section \uppercase\expandafter{\romannumeral5}.B.2) and different numbers of collected samples (section \uppercase\expandafter{\romannumeral5}.B.3). Then we evaluate how the designed core components contribute to performance improvements (section \uppercase\expandafter{\romannumeral5}.C). Furthermore, we construct synthetic datasets with different noise proportions to evaluate BinMLM in multi-programmer collaboration scenarios (section \uppercase\expandafter{\romannumeral5}.D). Finally, we explore the organization-level authorship verification ability of BinMLM on a real-world APT malware dataset (section \uppercase\expandafter{\romannumeral5}.E). 

\begin{figure*}[!t]
	\centering
	\includegraphics[width=14cm]{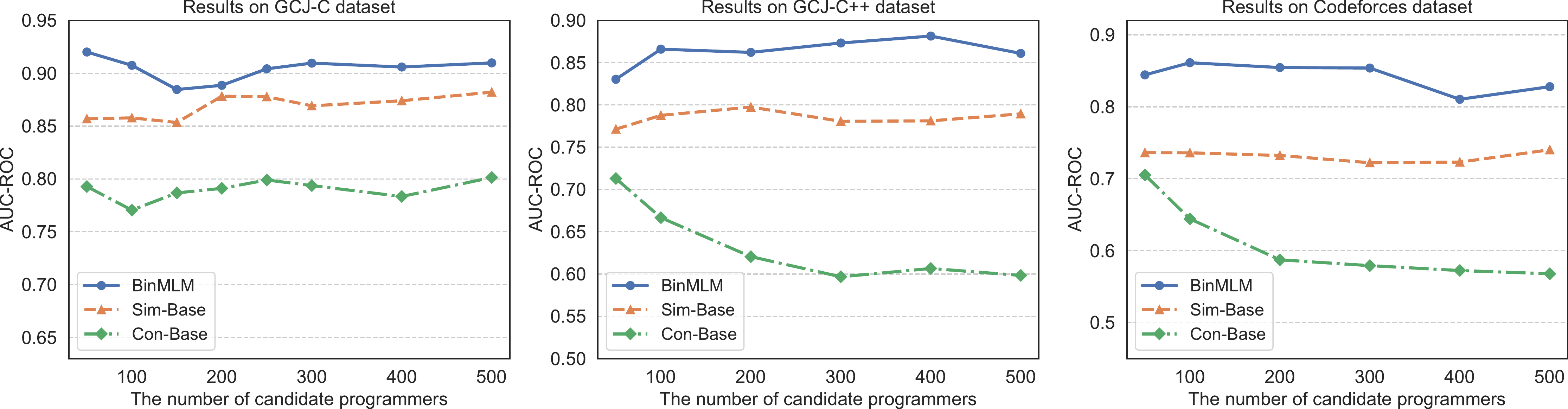}
	\caption{Results under under different numbers of programmers}
	\label{diff_programmers}
	\vspace*{-1.2\baselineskip}
\end{figure*}

\subsection{Experiment Setup}
\subsubsection{Datasets}
We use the following data sources for evaluation:
\textbf{(a)} Google Code Jam (GCJ): GCJ is an annual programming competition organized by Google \cite{gcj}. We collect participants' solutions written in C and C++ language from 2008 to 2020.
\textbf{(b)} Codeforces: We collect open submissions written in C++ from the competitive programming contest hosted by the online judge website Codeforces \cite{codeforces}. 
\textbf{(c)} APT-Malware: We collect malware samples of ten real-world APT groups from open source threat intelligence reports to evaluate BinMLM in security-related authorship verification scenarios.
	
For each data source, we construct the corresponding dataset according to the following rules: We randomly select $N$ programmers to construct the candidate set $P$. Each programmer has a small set of $m$ binaries as training samples with known annotations. In the verification phase, we sample anonymous binaries to construct authorship verification pairs. For a positive pair, we randomly select an unseen binary sample of the candidate programmer. For negative pairs, half of the test samples are developed by other $N-1$ programmers in the candidate set, and the others by randomly sampled remaining developers who do not belong to this set. The ratio of positive and negative verification pairs is 1:1.

\subsubsection{Evaluation metrics}
 We use two complementary metrics to evaluate BinMLM: AUC-ROC (Area under ROC curve) and AP (average precision). These two indicators can evaluate the performance of the binary authorship verification approaches without selecting specific thresholds.
 
 \subsubsection{Implementation Details}
 We implement our prototype with PyTorch framework and use dissembler $radare2$ \cite{radare2} to extract the  CFG and instruction sequences in the corresponding basic blocks. The parameters of our neural network is optimized by Adam \cite{kingma2014adam} with a learning rate of 1e-2, and we truncated the gradient of the language model for 20 time-steps. The hidden dimension of the embedding layer and LSTM encoder is set to 64. The number of the shared encoders is set to 5.  The weight of the regularization loss is set to 1e-4. Under this hyperparameter setting, BinMLM can achieve best results on our development sets.

\begin{figure*}[!t]
	\centering
	\includegraphics[width=14cm]{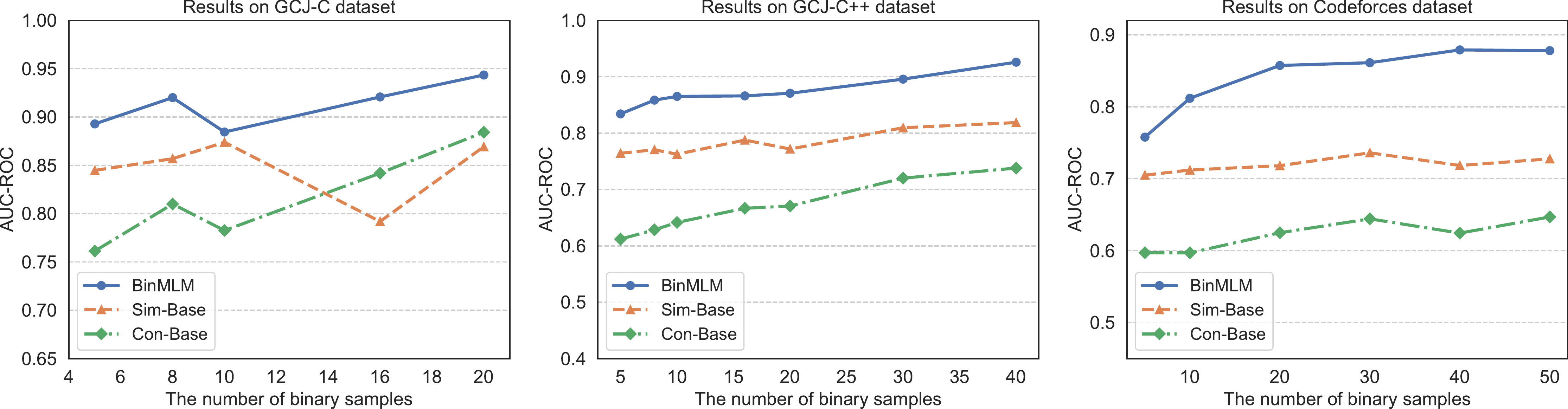}
	\caption{Results under different numbers of binary sapmles}
	\label{diff_samples}
	\vspace*{-1.2\baselineskip}
\end{figure*}

\subsection{Main results on different datasets}
\subsubsection{Comparision with Baselines}
Among the six existing binary authorship attribution approaches listed in table \ref{qualitative_sota_table}, only \cite{rosenblum2011wrote} and \cite{caliskancoding} can deal with programmers from the wild by stylistic similarity and classifier confidence score, respectively. We build our baselines based on these two implementation ways and conduct evaluations on the binary authorship verification task. We use the state-of-the-art feature set proposed by Caliskan-Islam \emph{et al.} \cite{caliskancoding}. Specifically, they extracted:
\textbf{(a)} Instruction traces of disassembly code, including instruction uni-grams, bi-grams, and tri-grams in a single line of assembly, and 6-grams spanning two consecutive assembly lines.
\textbf{(b)} Uni-grams and bi-grams of basic blocks extracted from the CFG of the binary sapmle.
\textbf{(c)} Abstract syntax tree (AST) node features extracted from the decompiled code.

Considering the time cost and the unstable success rate in real-world binary decompilation, we use the first two types of feature sets above, which can be easily constructed from disassembly code and shown to be important according to the feature selection results of the original paper.

We construct two groups of baselines based on the feature set. The first (\texttt{Sim-Base}) is built on similarity measures. For a known programmer $p_i$, we construct the feature vector of each collected sample and take the average as the programmer's stylistic representation. For an anonymous sample $b_j$, we build the feature vector in the same way and calculate the verification scores of the pair $\langle p_i, b_j \rangle$ by the cosine similarity between the corresponding vectors. We also use our proposed flow-aware opcode n-grams to construct variant feature sets.

The second group of baselines (\texttt{Con-Base}) is implemented based on the experiment designed by Caliskan-Islam \emph{et al.} \cite{caliskancoding} for the open-world scenario, in which the author of the test binary sample may not belong to the known programmers set. It exploits the confidence score of the classifier to determine whether to accept or reject the result. For a verification pair $\langle p_i, b_j \rangle$, we take the percentage of trees in the random forest that voted for $p_i$ when classifying $b_j$ as the confidence score, and then we use the normalized margin of the highest and second-highest confidences as the final verification score. Our implementation way is consistent with the original paper.

Table \ref{main_result_table} shows the binary authorship verification performance of \texttt{Sim-Base}, \texttt{Con-Base}, and BinMLM. We evaluate \texttt{Sim-Base} with different features combination ways and \texttt{Con-Base} with or without feature selection for comprehensive comparisons. Overall, the performance of \texttt{Sim-Base} is better than \texttt{Con-Base}, and BinMLM significantly surpasses them with 4.73\% $\sim$ 19.46\% improvement.

 \subsubsection{Performance under different numbers of programmers}
In this section, we evaluate the performance of BinMLM with the different candidate set sizes. Increasing the number of candidate programmers will test BinMLM's ability to characterize diverse developers' styles. Figure \ref{diff_programmers} shows the comparison of BinMLM and the two baselines \texttt{Con-Base} and \texttt{Sim-Base}. We gradually increase the number of candidates from 50 to 500 on GCJ-C, GCJ-C++, and Codeforces datasets. 

It can be seen that when the number of candidate programmers increases, the performance of \texttt{Con-Base} implemented by the classifier confidence score decreases significantly on GCJ-C++ and Codeforces datasets, while BinMLM and  \texttt{Sim-Base} remain relatively stable. When the number of programmers is in the range of 300 to 500, the AUC-ROC of \texttt{Con-Base} may be lower than 0.6, indicating a poor authorship verification ability. The AUC-ROC of \texttt{Sim-Base} ranges from 0.7231 to 0.8821. BinMLM significantly outperforms the two baselines under each setting. When the candidate set size is 500, the AUC-ROC values of BinMLM on the three datasets reach 0.9098, 0.8610, and 0.8279, respectively. This proves that our proposed author-specific programming style characterization method has strong versatility and can effectively perform the binary authorship verification for large-scale candidate programmers.

\subsubsection{Performance under different numbers of binary sapmles}
In real-world software forensics scenarios, especially security-related applications, the collected binaries are often very insufficient, and annotating the ground-truth is highly dependent on expert knowledge. Limited samples can not adequately reflect the author's programming behavior when implementing different functions. We evaluate BinMLM and the baselines with different numbers of support samples per candidate author. Similar to \cite{abuhamad2018large}, we merge the GCJ solutions with the same and complex participant ID from different years to increase the sample scales for comprehensive evaluations. Figure \ref{diff_samples} shows the comparison results. It can be seen that with the increase of the sample size, the performance of BinMLM has been steadily improved because the shared encoders and author-specific layers of the mixture-of-shared architecture can be trained better, and each developer's style can be extracted more accurately. However, more samples do not benefit \texttt{Con-Base} and \texttt{Sim-Base} as much because the diversified functions of each author's programs may lead \texttt{Con-Base}'s classifier to be confused and cause the similarity between the candidate author's profile and the test sample features unstable.

When the number of samples is very small, BinMLM still has significant advantages over \texttt{Con-Base} and \texttt{Sim-Base}. With only five support binaries, the AUC-ROC of BinMLM on the three datasets are 0.8929, 0.8341, and 0.7578, respectively. On Codeforces dataset with the more challenging setting of 100 authors and five support samples, the accuracy of BinMLM drops slightly, but still surpasses \texttt{Con-Base} and \texttt{Sim-Base}  by a large margin. It concludes that BinMLM can accurately extract the programming styles of candidate authors with very limited binary samples.  

\begin{table}
	\caption{Results of different opcode n-grams and different hops.}
	\label{opcode_ngram_table}
	\centering
	\begin{tabular}{p{100pt}<{\centering}|p{32pt}<{\centering}p{32pt}<{\centering}p{32pt}<{\centering}}
		\toprule
		Setting & GCJ-C & GCJ-C++ & Codeforces \\
		
		\midrule
		Opcode 1-gram & 0.8216 & 0.6935 & 0.7857 \\
		Opcode 2-gram & 0.8592 & 0.7529 & 0.8290\\
		Opcode 3-gram & 0.8671 & 0.8053 & 0.8519 \\
		Opcode 4-gram & 0.8816 & 0.8235 & 0.8561 \\
		
		\midrule
		Predict target units of 1-hop  & 0.8826 & 0.8204 & 0.8429 \\
		Predict target units of 2-hops & 0.8869 & 0.8285 & 0.8514 \\
		
		\midrule
		\textbf{BinMLM} (5-gram, 3-hops)& \textbf{0.8929} & \textbf{0.8414} & \textbf{0.8552}\\
		
		\bottomrule
	\end{tabular}
	
	\vspace*{-1.4\baselineskip}
\end{table}

\begin{table*}
	   \caption{Effectiveness of each component in BinMLM framework}
		\label{ablation_studies}
		\centering
		\begin{tabular}{p{120pt}<{\centering}|p{40pt}<{\centering}p{40pt}<{\centering}p{40pt}<{\centering}p{40pt}<{\centering}p{40pt}<{\centering}p{40pt}<{\centering}}
			\toprule
			
			\multirow{2}{*}{Approach} & \multicolumn{2}{c}{GCJ-C} & \multicolumn{2}{c}{GCJ-C++} & \multicolumn{2}{c}{Codeforces}  \\
			
			& AUC-ROC & AP & AUC-ROC & AP & AUC-ROC & AP  \\
			
			\midrule
			\emph{MoS + OPT + REG} (\textbf{BinMLM}) & \textbf{0.8929} & \textbf{0.9135} & \textbf{0.8414} & \textbf{0.8588} & \textbf{0.8552} & \textbf{0.8640} \\
			
			\emph{MoS + OPT} & 0.8632 & 0.8856 & 0.7682 & 0.7670 & 0.8451 & 0.8523 \\
			
			\emph{MoS} & 0.8703 & 0.8901 & 0.7682 & 0.7666 & 0.8421 & 0.8495 \\
			
			\emph{Single-encoder} & 0.8578 & 0.8809 & 0.7528 & 0.7483 & 0.8380 & 0.8453 \\
			
			\emph{Naive} & 0.8492 & 0.8760 & 0.7357 & 0.7397 & 0.8273 & 0.8309 \\
			
			\bottomrule
		\end{tabular}
	\vspace*{-1.0\baselineskip}
\end{table*}

\begin{table*}
	\caption{Results on multi-programmer collaboration datasets}
		\label{multiauthor_res_table}
		\centering
		\begin{tabular}{p{55pt}<{\centering}|p{38pt}<{\centering}p{38pt}<{\centering}p{38pt}<{\centering}|p{38pt}<{\centering}p{38pt}<{\centering}p{38pt}<{\centering}|p{38pt}<{\centering}p{38pt}<{\centering}p{38pt}<{\centering}}
			\toprule
			
			\multirow{2}{*}{Major Proportion} & \multicolumn{3}{c|}{GCJ-C (N = 50, m = 8)} & \multicolumn{3}{c|}{GCJ-C++ (N = 100, m = 10)} & \multicolumn{3}{c}{Codeforces (N = 100, m = 30)}  \\
			
			& \texttt{Con-Base} & \texttt{Sim-Base} & BinMLM & \texttt{Con-Base} & \texttt{Sim-Base} & BinMLM & \texttt{Con-Base} & \texttt{Sim-Base} & BinMLM \\
			\midrule
			
			1.0 & 0.8100 & 0.8569 & \textbf{0.9201} & 0.6416 & 0.7779 & \textbf{0.8652} & 0.6442 & 0.7360 & \textbf{0.8612} \\
			0.9 & 0.7786 & 0.8497 & \textbf{0.9020} & 0.6347 & 0.7778 & \textbf{0.8419} & 0.6241 & 0.6347 & \textbf{0.8376} \\
			0.8 & 0.7769 & 0.8621 & \textbf{0.9088} & 0.6063 & 0.7642 & \textbf{0.8344} & 0.6083 & 0.7090 & \textbf{0.8310} \\
			0.7 & 0.7677 & 0.8264 & \textbf{0.9049} & 0.6013 & 0.7628 & \textbf{0.8210} & 0.6170 & 0.6938 & \textbf{0.8324} \\
			0.6 & 0.7455 & 0.7949 & \textbf{0.8736} &  0.6051 & 0.7549 & \textbf{0.8346} & 0.5972 & 0.6873 & \textbf{0.8048} \\
			
			\bottomrule
		\end{tabular}
	\vspace*{-1.4\baselineskip}
\end{table*}

\subsection{Ablation studies of BinMLM}
We design three groups of ablation studies to evaluate the effects of BinMLM's core components. The dataset settings of our ablation studies are the same as section V.B.1.

The first group of ablation studies compare opcode n-grams with different values of $n$, and results in table \ref{opcode_ngram_table} show that the performance is positively correlated with the value of $n$ in the range of 1 to 5. When the $n$ exceeds 5, we still observe a slight improvement, but we do not increase it further due to limited time and computing resources. 

The second group of ablation studies evaluate the performance when predicting subsequent units of different hops. As shown in table \ref{opcode_ngram_table}, predicting target units of more hops can improve the verification performance. We finally set $n$ to 5 and the number of hops to 3. 

The third group of ablation studies evaluate the contribution of
each part of our model. We set up five variant models for comparison: \textbf{(a)} \emph{MoS (mixture-of-shared architecture) + OPT + REG} (original BinMLM): The \emph{MoS} architecture of BinMLM contains multiple shared encoders to extract common programming patterns from different views, and a separate gate layer for each programmer to learn the combination weights of shared representations. We train \emph{MoS} with the optimization pipeline combining joint-training and author-specific fine-tuning. Then we add the regularization loss in the fine-tuning stage to encourage the parameters of specific linear decoders to be more similar and prevent overfitting.
\textbf{(b)} \emph{MoS + OPT}: Remove the regularization loss in the fine-tuning stage.
\textbf{(c)} \emph{MoS}: Remove the optimization pipeline of BinMLM. 
\textbf{(d)} \emph{Single-encoder}: Set up a single RNN encoder shared among all programmers.
\textbf{(e)} \emph{Naive} architecture: Train a separate RNN language model for each programmer. 

Table \ref{ablation_studies} shows the performance of BinMLM and its variant models on the three datasets. The results prove that the mixture-of-shared architecture, optimization pipeline, and specially designed regularization loss significantly improve BinMLM on the binary authorship verification task.

\subsection{Robutness on multi-programmer collaboration datasets}
Modern software is usually developed through cooperation. Code snippets of collaborators will introduce noise when characterizing the major developer's style. We construct synthetic datasets to evaluate BinMLM in multi-programmer collaboration scenarios. Gong \emph{et al.} \cite{gong2021code} conducted an empirical study on open source software projects. They found that for most program files, about 80\% of their code lines are contributed by a single programmer. So we assume each program is completed by a major contributor and several other developers. For a verification pair $\langle p_i, b_j \rangle$, we randomly select two other programmers as collaborators and mix their program fragments into the sample $b_j$. We adjust the proportion of the collaborator's fragments to control the difficulty of the task, and we ensure that the proportion of the program developed by the original author of $b_j$ remains above 60\%.

Table \ref{multiauthor_res_table} shows the results on the multi-programmer collaborative datasets. With the increase of the proportion of collaborators' programs, the difficulty of modeling the style of the major developer has increased, and the performance of the three approaches has decreased correspondingly. But under each setting, BinMLM still significantly outperforms the baselines based on similarity metrics and classifier confidence scores. The AUC-RO of BinMLM remains above 0.8, and the decline degree ranges from 0.0306 to 0.0564. It proves that our designed programmer's style characterizing method based on the mixture-of-shared language models and specific optimization pipeline performs more robust when dealing with the mixed programming style of collaborators.

\subsection{Case study in a real-world scenario}
In this section, we provide case study to demonstrate the effectiveness of BinMLM in real-world scenarios. Binary analysts care about the common coding habits of authors in a specific organization, but existing work mainly conducts manual analysis of individual cases \cite{marquis2015big} \cite{alrabaee2018leveraging}.
We perform systematic experiments on the authorship verification of malware samples produced by specific APT (Advanced Persistent Threat) attack groups. APT attacks target particular companies or nations for political or commercial motivations. The malware samples adopt in the attack are likely to contain the identity information of the hacker team. 

We collect malware from public threat intelligence sources and construct the APT-Malware dataset. The malware comes from ten famous APT events includes \emph{Equation Group}, \emph{Gorgon Group}, \emph{DarkHotel}, \emph{Energetic Bear}, \emph{Winnti}, \emph{APT 10}, \emph{APT 21}, \emph{APT 28}, \emph{APT 29}, and \emph{APT 30}. Compared with the GCJ and Codeforces datasets, the binary samples of the APT-Malware dataset are much more complicated, and the label granularity is coarser, reflecting that the samples were developed by multiple collaborators in specific organizations. 

To perform staged attacks, malware belonging to the same APT group may implement very different functions, and some samples are intentionally obfuscated to prevent detection by anti-virus products, which makes it more difficult to locate the attacker's identity. We randomly select five support samples for each attack group to train the APT-specific language models. Considering data imbalance in real-world scenarios, where the proportion of negative verification pairs should be relatively larger, we set the ratio of the negative pairs to positive pairs in the range of 1 to 8. Half of the test samples in negative pairs are from the benign developers of the GCJ dataset.

\begin{figure}[!t]
	\centering
	\includegraphics[width=7.7cm]{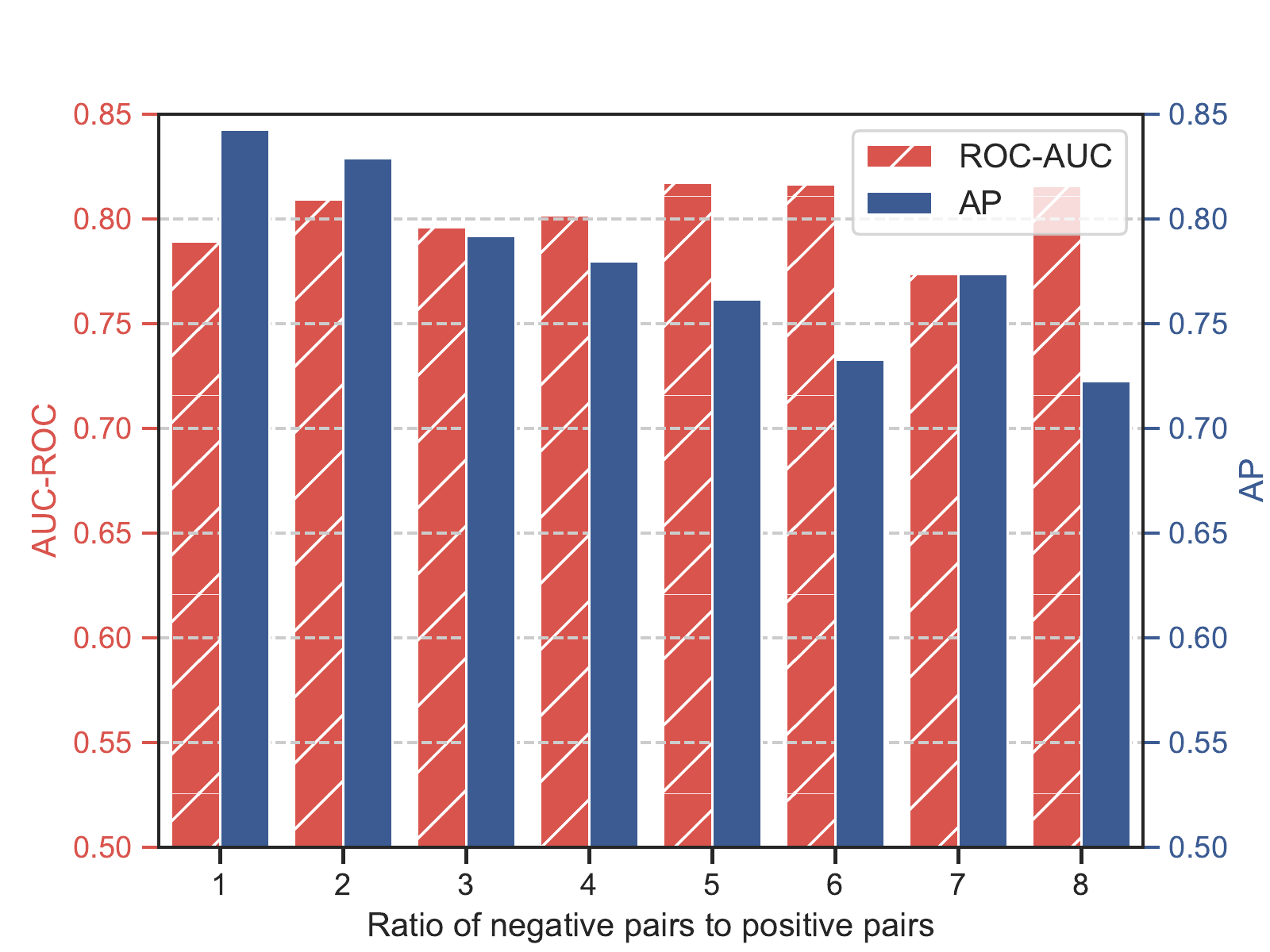}
	\caption{Results on the APT-Malware dataset}
	\label{apt_malware_res}
\vspace*{-1.4\baselineskip}
\end{figure}

Figure \ref{apt_malware_res} shows the authorship verification performance of BinMLM on the APT-Malware dataset under different negative to positive ratios. We can see that the AUC-ROC value of BinMLM is relatively stable and remains in the range of 0.7736 to 0.8172, while as the proportion of negative verification pairs increases, the AP value gradually decreases since it is more sensitive to the skewed datasets. When the ratio of the positive and negative pairs is 1:5, the AUC-ROC and AP of BinMLM are 0.8172 and 0.7617, respectively. Overall, BinMLM can extract programming style information related to identity characteristics from APT groups' malware samples and achieve organization-level authorship verification with decent performance, providing valuable auxiliary evidence for identifying the organization behind the APT event.

\section{Related Work}
\subsection{Source code authorship analysis}
Source code contains rich style characteristics which can be very helpful in software forensics, such as plagiarism detection and copyright investigation \cite{yang2017authorship} \cite{burrows2014comparing} \cite{caliskan2015anonymizing} \cite{alsulami2017source} \cite{kang2019assessing} \cite{bogomolov2021authorship} \cite{gong2021code}  \cite{wang20203}. For source code de-anonymizing, Caliskan-Islam \emph{et al.} \cite{caliskan2015anonymizing} constructed a dynamic code stylometry feature set including lexical, layout, and syntactic features. Abuhamad \emph{et al.} \cite{abuhamad2018large} selected code n-grams with high TF-IDF scores and fed them into the RNN network to extract the deep representations. Alsulami \emph{et al.} \cite{alsulami2017source} converted the AST into multiple subtrees and applied hierarchical bi-LSTM to extract the syntax features. Bogomolov \emph{et al.} \cite{bogomolov2021authorship} designed two language-agnostic models work with path-based code fragment representations. Some researchers performed authorship identification on Android platform \cite{kalgutkar2018android} \cite{kalgutkar2018android2} \cite{wang20203}. Wang \cite{wang20203} divided packages of Android applications into modules and built the author's coding habit fingerprint of the main module.

 \subsection{Binary authorship analysis}
 Authorship analysis targeting binaries is significant, because analysts cannot obtain source code in many real-world scenarios \cite{marquis2015big} \cite{rosenblum2011wrote}   \cite{alrabaee2014oba2} \cite{alrabaee2018leveraging} \cite{alrabaee2019bineye}. Rosenblum \emph{et al.} \cite{rosenblum2011wrote} built author stylistic templates on features like instruction idioms and CFG graphlets. Alrabaee \emph{et al.} \cite{alrabaee2018leveraging} first filtered out irrelevant compiler functions, then characterized the author's habits contained in user-defined functions by programming choices. Caliskan-Islam \emph{et al.} \cite{caliskancoding}  extracted multi-source style features from disassembly code and the AST of decompiled code. Meng \emph{et al.} \cite{meng2016fine} conducted empirical analysis on open source software and concluded that binary authorship attribution should be performed at the basic block granularity. The manual feature engineering processes of the above approaches are dataset-dependent and rely on domain knowledge. BinEye \cite{alrabaee2019bineye} proposed an intelligent deep learning-based binary authorship attribution model. It applied three CNNs on binary gray-scale images, API call sequence, and opcode sequences. 

 \subsection{Text authorship analysis}
 Text authorship analysis has a more extended research history \cite{bagnall2015author} \cite{ouyang2020gated} \cite{kalgutkar2019code} \cite{ding2017learning} \cite{bevendorff2019bias}. Our task setting is inspired by the open-world text stylometry problems defined by Stolerman \emph{et al.} \cite{stolerman2013classify} and the shared PAN series of competitions \cite{pannetwork}. For text authorship verification, Bagnall \cite{bagnall2015author} designed a char-level language model with a single encoder and independent softmax groups to model the authors' writing styles, which won first place in the PAN-2015 competition.  Ouyang \emph{et al.} \cite{ouyang2020gated} found that the POS (Part of Speech)-level language model can better characterize authors' syntactic styles. They designed a gated unit in the decoding stage to integrate common writing patterns and specific styles. 
 
\section{Conclusion}
In this paper, we formulate a practical binary authorship verification task, which can handle binary samples from unknown programmers and accurately reflect the real-life experiences of software forensic experts. We implement a binary authorship verification framework, BinMLM. It trains author-specific language models on the flow-aware opcode n-gram sequences to automatically characterize the developer's programming styles. We exploit a mixture-of-shared architecture to fully use the limit training samples and model the developer's combination preference of multiple universal programming patterns. Through an effective optimization pipeline, BinMLM can separate the programmer's unique style from a large portion of general patterns. Extensive experiments show that BinMLM outperforms baselines built on the state-of-the-art feature set by a large margin and remains robust in multi-author collaboration scenarios and organization-level verification on real-world APT malware datasets.

\section*{Acknowledgment}
The authors would like to thank the anonymous reviewers for their insightful comments. This work was supported by the National Natural Science Foundation of China under Grant U1736218. The corresponding author is Yongzheng Zhang.

\small
\bibliographystyle{IEEEtran}
\bibliography{cameraready_BinMLM}

\end{document}